\documentclass[10pt,twocolumn]{wlscirep}
\usepackage[utf8]{inputenc}
\usepackage[T1]{fontenc}
\usepackage{graphicx}
\usepackage{amsmath}
\usepackage{amssymb}
\usepackage{epstopdf}
\usepackage{subfigure}
\usepackage{bm}
\usepackage{dsfont}
\usepackage{hyperref}
\usepackage{braket}
\usepackage{todonotes}
\usepackage{physics}
\usepackage{comment}
\usepackage{physics}
\usepackage[outercaption]{sidecap}    
\usepackage[mathscr]{eucal}

\newcommand{\mi}{\mathrm{i}}

\usepackage{cleveref}

\title{Studying the effect of  fluctuating environment on intra-atomic frequency comb based quantum memory}

\author[1,*]{G.~P.~Teja}
\author[1]{Sandeep K.~Goyal}

\affil[1]{Indian Institute of Science Education and research, Mohali, Punjab 140306, India}

\affil[*]{teja4477@gmail.com}

\begin{document}

\begin{abstract}
  In this article, we study the effect of various environmental factors on intra-atomic frequency comb (I-AFC) based quantum memory. The effect of the environment is incorporated as random fluctuations and non-uniformity in the parameters such as comb spacing and the optical depth, of the frequency comb. 
  We found that the I-AFC is viable for photon storage even for very large fluctuations in the parameters of the frequency comb, which makes I-AFC a robust platform for photon storage. Furthermore, we show that the non-uniform frequency combs without any fluctuations in the comb parameters can also yield efficient quantum memory. Since the intra-atomic frequency combs found in natural atomic systems are often non-uniform, our results suggest that   a large class of these systems can be used for I-AFC based efficient quantum memory.
\end{abstract}

\flushbottom
\maketitle
%
%
\thispagestyle{empty}


\section{Introduction}

Quantum memory is a device which can store and reemit photons on demand. Quantum memory is essential for photonic quantum information processing and long-distance quantum communications~\cite{duan2001,sango2011,zhang2015}. Along with probabilistic single-photon sources, it can also be used to achieve deterministic single-photon sources~\cite{chen2006,barros2009}. Among the atomic ensemble based quantum memories electromagnetically induced transparency~\cite{fleischhauer2002,julsgaard2004,gorshkov2007,cho2010,hockel2010,wang19,jiang2019}, controlled reversible inhomogeneous broadening~\cite{Nilsson2005,Alexander2006,Kraus2006,San2007,Longdell2008,Iakoupov2013}, gradient echo memory~\cite{hetet08,sparkes2010,hosseini11,hosseini2012}, Raman quantum memory~\cite{Reim2010,thomas2017,guo19,kalachev2019} and 
the atomic frequency combs (AFCs)~\cite{de2008,afzelius09,lvovsky2009,usmani2010,bonarota2011,Jobez2016,teja19} are the most prevalent protocols for photonic quantum memory. The basic idea behind an atomic ensemble based quantum memory is the controlled reversible transfer of information between the light field and the atomic states. The incoming photons are made to interact with the ensemble of atoms and the excitation is transferred to a long-lived state. Reversing the process results in the photon emission.

Atomic frequency comb based quantum memory relies in artificially created frequency  comb by reshaping the  inhomogeneously broadened spectrum by means of optical hole burning in an ensemble of atoms. The incoming photon is absorbed as a delocalized excitation over the frequency comb. The comb-like structure of the atomic spectrum results in a photon-echo at a later time; hence the frequency comb serves as a delay line for the photon. To achieve a on-demand quantum memory, the excitation can be transferred to a long-lived spin state by applying an appropriate $\pi$-pulse. By applying another $\pi$-pulse the excitation can be transferred back to the excited state which will be emitted in the photon-echo.

In the AFC based quantum memory, the photons are stored as a delocalized excitation over all the teeth of the frequency comb which consist of billions of atoms. Therefore, in order for AFC to work the entire atomic ensemble must behave like a single quantum system. A small relative fluctuation in the frequencies of different atoms may give rise to strong decoherence in the frequency comb resulting in no photon-echo. This restricts the temperature range of AFC based quantum memory to a few Kelvins.

The intra-atomic frequency comb (I-AFC) based quantum memory is operationally similar to the one using AFC~\cite{teja19}. The difference being that the frequency comb in I-AFC is constructed using the degenerate hyperfine energy levels of an atom. The degeneracy in the hyperfine levels is lifted by applying external magnetic field. However, there are certain limitations of I-AFC which may affect the efficiency of quantum memory.

One of the limitation of the I-AFC is that the frequency combs obtained in natural atomic systems are not always uniform. The non-uniformity can be due to unequal absorption or due to unequal spacing between  different hyperfine states. This may severely affect the storage efficiency, forcing us to choose the atomic systems which have almost uniform frequency combs. Further, the local fluctuations in the electromagnetic field can cause fluctuations in various parameters of the frequency comb. Both of these internal and external factors can cause low-efficiency for photon storage. In this article, we numerically study the effect of all these adversities on the efficiency of the I-AFC based quantum memory. Effect of random fluctuations in the absorption and the comb spacing is incorporated by introducing randomness in the said parameters stochastically and then the macroscopic polarization is obtained by averaging over the fluctuations.

Our study shows that the fluctuations in different parameters in the I-AFC affect the efficiency of the photon storage differently. For example, the fluctuations in the absorption in different teeth of the comb, i.e., non-uniformity in the height of the teeth has negligible effect on the efficiency. Whereas the fluctuations in the comb spacing has significant effect. Fortunately, this adverse effect can be  mitigated by increasing the finesse of the frequency comb, which can be done by increasing the external magnetic field. The non-uniformity of the frequency comb has little effect on the efficiency of the quantum memory. Therefore, our study conclusively establish that a large class of atomic systems can be used as photonic quantum memory using I-AFC protocol.

The article is organized as follows: in Sec.~\ref{Sec:Background} we introduce the concepts useful to understand our results and calculations. Here we briefly explain the concept of AFC and I-AFC based quantum memory and the calculations for photon-echo efficiencies in the forward and backward propagation. In Sec.~\ref{Sec:Results} we present our results  and numerical simulations for the effect of random environment on the efficiency in I-AFC.  We conclude in Sec.~\ref{Sec:Conclusion}.

\section{Background}\label{Sec:Background}
In this section, we discuss the topics which are relevant for results presented in this article. We start by discussing the AFC protocol for quantum memory followed by I-AFC protocol.

\subsection{AFC}
\begin{figure}[!htb]
\subfigure[\label{com1}]{\includegraphics[width=4cm,height=3cm]{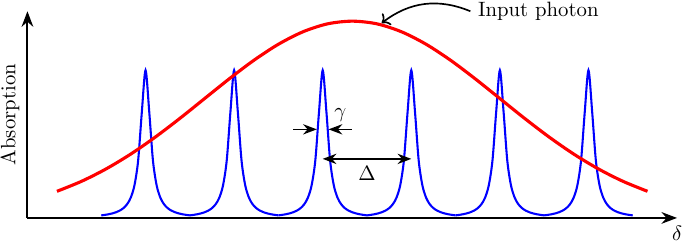}}
\subfigure[\label{eff}]{\includegraphics[scale=0.4]{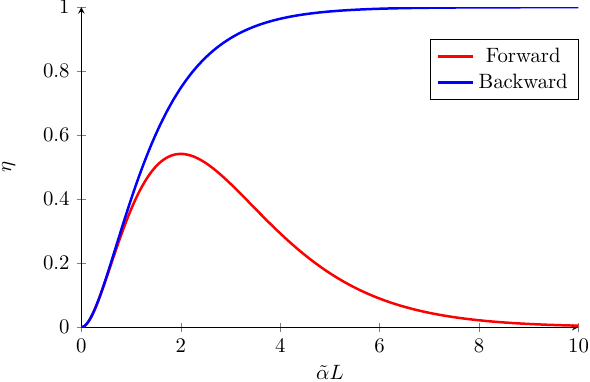}}
\caption{ \subref{com1} AFC with tooth width $\gamma$ and comb spacing $\Delta$. The red curve shows the spectrum of width $\gamma_p$ of the incoming photon which interacts with the AFC. For the AFC to work efficiently $\gamma_p \gg \Delta$. \subref{eff} Forward and backward efficiency $\eta$ for AFC as a function of optical depth $\tilde \alpha L$.}
\end{figure}

AFC typically consists of rare-earth ions doped in a dielectric crystal  \cite{de2008,afzelius09,usmani2010,bonarota2011,lauritzen2011,timoney2012,zhou2013}. 
The rare-earth ions when doped in crystals experience in-homogeneously broadened spectral lines due to their interaction with the local environment in the host material. This spectrum can be reshaped by applying narrow-band lasers to transfer a fraction of population of ions corresponding to a chosen frequency to a stable auxiliary state, resulting in a hole in the spectrum. Repeating the process at desired frequencies reshapes the  spectrum to yield a comb like structure  [Fig.~\ref{com1}]. In a uniform AFC, the spacing between the neighboring teeth of the comb is fixed (say $\Delta$) and every tooth has a width of $\gamma$. In the approximation that $\gamma \ll \Delta$ the frequency of the $n$-th tooth can be written as $\delta_n = \omega_0 + n\Delta$ around some mean frequency $\omega_0$. The total size of the frequency comb is $\Gamma = N\Delta \gg \Delta $ for $N$ number of teeth.

When a photon of spectral width $\gamma_p \gg \Delta$ is absorbed in the AFC with $N$ number of teeth, the state of the AFC can be formally written as~\cite{dicke1854,gross1982}
\begin{align}
\Ket{\Psi}_{\text{AFC}} &= \sum_{j=1}^N \left(c_je^{\mi\delta_j t}\ket{\{e_j\}} \prod_{k\ne j} \ket{\{g_k\}}\right).\label{Eq:AFC-Dicke}
\end{align} 
Here $\ket{\{g_j\}}$  and $\ket{\{e_j\}}$ represent the ground and collective single-excitation state of all the atoms with detuning $\delta_j$, respectively, and the $c_j$'s represent the absorption coefficient of each tooth in the comb.
 The probability of photon emission from the AFC is  $P(t) = |\sum_{j=1}^{N} c_j e^{\mi j t \Delta }|^2$ where we have used $\delta_j = j\Delta$. The $P(t)$ vanishes at all times $t$ except for $t_n =  2n\pi/\Delta$ for positive integers $n$ resulting in $n$-th photon-echo.

The collective dynamics of electric field and the atomic state of the AFC  in the weak  field approximation is governed by the following Maxwell-Bloch equations~\cite{sangouard2007,gorshkov2007}
\begin{align}
\pdv{\rho_{eg}}{t}&= (-\mi \delta-\dfrac{\gamma}{2}) \rho_{eg} + \mi \dfrac{d_{eg}}{2\hslash} \mathcal{E}(z,t), \label{mb1}\\
\qty(\pdv{t}\pm c\pdv{z} )\mathcal{E}(z,t)&= \mi \dfrac{\omega_L d_{ge}}{ \epsilon_0 V} \int n(\delta) \rho_{eg}(z,t,\delta) \,\text{d}\delta. \label{mb2}
\end{align}
Here $n(\delta)$ is the atomic spectral distribution which characterize the AFC,  $\delta=\omega_{eg}-\omega_L$ is the detuning between light $\omega_L$ and atomic transition $\omega_{eg}$. The $\rho_{eg} \equiv \rho_{eg}(z,t)$ is off-diagonal density matrix element, $d_{eg}$ is the transition dipole matrix element for the transition $\ket{e}$-$\ket{g}$. The $\pm$ sign in Eq.~\eqref{mb2} represent the forward and backward propagating modes of light. For the case of forward propagating modes, solving these equations yields the output electric field $\mathcal{E}_f$ as a function of $z$ and the frequency $\omega$, which reads
\begin{align}
\mathcal{E}_f(z,\omega)= \mathcal{E}_f(0,\omega)e^{-\mathcal{D}z}.
\end{align}
The input and the output electric field are related by the propagator  $\mathcal{D}$ which is given by
\begin{align}
\mathcal{D} =\alpha \int \dfrac{n(\delta)}{\mi(\omega + \delta)+\gamma/2} \,\text{d}\delta +\dfrac{\mi \omega}{c},&& 
\alpha = \dfrac{\abs{d_{eg}}^{2}\omega_{L}}{2\hslash\epsilon_0 c V}.\label{Eq:D_AFC}
\end{align}
Here $\alpha$ is the absorption coefficient and $V$ is the volume of the atomic ensemble.

In Principle, the electric field in the time domain $\mathcal{E}_f(z,t)$ can be calculated by taking the inverse Fourier transform of $\mathcal{E}_f(z,\omega)$; however, the expression for the same will be very cumbersome. In the limit $\Gamma\gg\gamma_p\gg\Delta$,  and after propagating the electric field for a distance $L$,  a simplified  expression for $\mathcal{E}_f(L,t)$ can be written as~\cite{afzelius09}
\begin{align}
\mathcal{E}_f(L,t)&= e^{- \qty(\sqrt{2}\pi/\mathcal{F})^2} \left(\tilde{\alpha} L e^{-\tilde{\alpha} L/2}\right) \, \mathcal{E}_f(0,t-2\pi/\Delta), \label{fme}
\end{align}
with  $\tilde{\alpha}=\alpha/\mathcal{F}$ and the finesse $\mathcal{F}$ of AFC is the ratio between comb spacing and tooth width, i.e., $\mathcal{F}=\Delta/\gamma$. 

 The AFC on its own results in a photon-echo which can be thought of as a delay line which is typically of the order of  microseconds and solely depends on the comb spacing $\Delta$. In order to achieve on-demand retrieval of the input photon the excitation is transferred from the $\ket{e}$ state to a long-lived spin state $\ket{s}$ by applying a $\pi$-pulse. After the storage time (limited by the lifetime of the spin state $\ket{s}$) a second $\pi$-pulse is applied to transfer the excitation back to $\ket{e}$ which due to AFC will re-phase after time $2\pi/\Delta$ \cite{usmani2010}. Thus, one can achieve an on-demand and deterministic quantum memory.

 Application of two $\pi$-pulses results in the overall sign change in the electric field which causes backward propagation of light. If we apply the  $\pi$-pulses  at time  $t=\pi/\Delta$ the atomic polarization induced by the input electric field will 
 act as the source term for the backward field propagation. On solving the Maxwell-Bloch equations, the electric field in the backward mode can be written as~\cite{afzelius09}
\begin{align}
\mathcal{E}_b(L,t)&=  e^{- \qty(\sqrt{2}\pi/\mathcal{F})^2} (1-e^{-\tilde{\alpha} L})\,\mathcal{E}_f(0,t-2\pi/\Delta),\label{bms}
\end{align}
which is same as forward mode solution apart from the factor $1-e^{-\tilde{\alpha} L}$ instead of $\tilde{\alpha} L e^{-\tilde{\alpha} L/2}$.

The quality of the quantum memory is quantified by the efficiency $\eta$ and the fidelity $\mathfrak{F}$. The efficiency is defined as the ratio of the intensity of light in the first echo to the input light and the temporal fidelity is defined as the overlap between the input pulse and and the first echo i.e.,
\begin{align}
\eta &= \dfrac{\int_{\pi/\Delta}^{3\pi/\Delta} \abs{\mathcal{E}(L,t)}^2 \, dt }{ \int \abs{\mathcal{E}(0,t)}^2\, dt}.\\
\mathfrak{F}&= \dfrac{\abs{\int_{\pi/\Delta}^{3\pi/\Delta} \mathcal{E}^*(0,t-{2\pi}/{\Delta}) ~\mathcal{E}(L,t) \, dt}^2 }{ \int \abs{\mathcal{E}(0,t-{2\pi}/{\Delta})}^2\, dt ~ \int_{\pi/\Delta}^{3\pi/\Delta} \abs{\mathcal{E}(L,t)}^2  \, dt }. \label{fmo}
\end{align}  
where the input is shifted by the first echo time $2\pi/\Delta$  for the fidelity.

 For high finesse ($\mathcal{F} \gg 1$) the efficiency for the forward propagating modes of light becomes $\eta_f = (\tilde{\alpha} L)^2 \exp(-\tilde{\alpha} L)$ which can approach to a maximum value of $54\%$ for $\tilde{\alpha}L\equiv \alpha L/\mathcal{F}=2$ (Fig.~\ref{eff}). Since $\mathcal{F}\gg 1$ the forward mode requires high absorption ($\alpha L > \mathcal{F}$) for maximum efficiency.  The efficiency in the backward-mode is 
$\eta_b(L) = (1-e^{-\tilde{\alpha} L})^2$
which can be optimized over absorption and finesse to reach $100\%$ (Fig.~\ref{eff}).

In order for AFC to work as a quantum memory, a large number the atoms in the atomic ensemble need to work as a single macroscopic quantum system. The incoming photon is absorbed as a collective excitation over all the atoms coherently. Any small phase fluctuation between these atoms will cause the AFC to fail. This makes the AFC protocol for quantum memory very sensitive to local environment.  To overcome this problem one can  exploit the degeneracy in the atomic states of alkali atoms to realize a frequency comb. Such frequency combs are called I-AFC, which is introduced in the following subsection.

\subsection{I-AFC}\label{siafc}
In I-AFC we start by considering an optical transition between hyper-fine degenerate energy levels $\{\ket{g_m}\}$ and $\{\ket{e_n}\}$ of an alkali atom. The degeneracy in the excited and ground states is lifted by applying external magnetic field with splitting proportional to magnetic field. Collectively, all the dipole allowed transitions between the ground state manifold and the excited state manifold yield a comb like structure similar to the one in Fig.~\ref{com1}, which is known as I-AFC \cite{teja19}.

The propagation of electromagnetic field through an ensemble of atom possessing I-AFC can be calculated using the Maxwell-Bloch equations. The propagator $\mathcal{D}$ and absorption coefficient $\alpha$ for the dynamics reads
\begin{align}
\mathcal{D} = \sum_{nm} \dfrac{\alpha_{nm}}{1/2 + \mi/\gamma (n \Delta+\omega)}+\dfrac{\mi \omega}{c},\label{Eq:D_IAFC} \qquad
\alpha_{nm} =\mathcal{N} \rho_{mm} \dfrac{\abs{d_{nm}}^{2}\omega_{L}}{2\hslash \epsilon_0 c \gamma}.
\end{align}
Here $\mathcal{E}(0,\omega)$ is the input pulse with the mean frequency $\omega_L$, and  $\gamma$, $\mathcal{N}$ and $d_{nm}$ are the  tooth width, number density of atoms and the transition dipole moment between $n$th excited and $m$th ground state, respectively.

Similar to AFC, the dynamics of the electric field in I-AFC is completely characterized by the propagator $\mathcal{D}$, which inturn is controlled by the finesse $\mathcal{F}$ and absorption coefficient $\alpha$. Despite the differences between  the propagators of AFC and I-AFC [Eqs.~\eqref{Eq:D_AFC} and \eqref{Eq:D_IAFC}], they yield similar results for photon-echo and efficiencies favoring high finesse and optical depths~\cite{teja19}.

The calculations for the forward efficiency $\eta_f$ in I-AFC can be done by calculating the ratio between the  intensities in first echo and the total input intensity, as given in Eq.~\eqref{fmo}. However, the efficiency for the back scattering $\eta_b$ requires an indirect approach~\cite{teja19}. In order to calculate $\eta_b$, first we estimate the average $\tilde\alpha$ for I-AFC by comparing the  I-AFC data with the AFC data for the forward propagation. By comparing forward efficiencies of I-AFC with the Eq.~(\ref{fme}) we obtain the common overall factor ($ e^{- \qty(\sqrt{2}\pi/\mathcal{F})^2}$) which along with $\tilde{\alpha}$ is used to calculate the backward efficiencies.

If the atomic ensemble is at temperature $T$, then the effect of temperature can be incorporated  by adding Doppler shifts $\vb{k\cdot v}$ to detunings $\delta$ i.e. $\delta \to \delta-\vb{k.v}$ where $\vb{v}$ is the thermal velocity of the atoms and $\vb{k}$ is the wavevector of the incoming light. Then the macroscopic polarization is obtained by taking average over the entire velocity range with the corresponding probability distribution $P(\vb{v}) \propto \exp(-mv^2/2k_bT)$, i.e., 
\begin{align}
  \mathcal{P}( \omega) & = {\mathcal E}( \omega) \int  \mathcal{D}( \omega,\vb{v})  ~P(\vb{v}) ~ d\vb{v}.
\end{align}
Here $k_b$ is the Boltzman constant and $m$ is the mass of the atom. 
For simplicity we approximate the Gaussian probability distribution with a Lorentzian distribution~\cite{teja19} to obtain an analytic expression for the propagator $\mathcal{D}( \omega)$ as
\begin{align}
  \mathcal{D}( \omega) & =  \sum_{n} \dfrac{\gamma \alpha_n}{\gamma/2+\Gamma_T+\mi(\delta_{n}+\omega)}+\dfrac{\mi \omega}{c},\label{ap}
\end{align}
where $\Gamma_T$ is the Doppler broadening at temperature $T$ and the finesse $\mathcal{F}$ is redefined with Doppler width as  $\mathcal{F}=\dfrac{\Delta}{\gamma+2 \Gamma_T}$.  For the rest of the article we set  $T=10$K and finally Eq.~\eqref{ap} results in the  output  electric~\cite{teja19}
\begin{align}
\mathcal{E}(L,\omega)= \mathcal{E}(0,\omega)e^{-\mathcal{D}L}.\label{iafc}
\end{align}

For simplicity we have assumed here that the spacing between the teeth of the comb $\Delta$ is uniform. In real system such as Cesium atoms, the comb spacing (controlled by external magnetic field) and the absorption coefficient of each of the tooth can be non-uniform. Furthermore, local fluctuations in the electric or magnetic fields can give rise to fluctuations in  the teeth spacing and variation in the height of each teeth  of the frequency comb. These factors can result in non-optimum efficiency of the quantum memory. In the subsequent section, we study the effect of non-uniformity and local fluctuations on the efficiency of I-AFC based quantum memory. We also give results for the randomness in IAFC Cesium atom.

\section{Results}\label{Sec:Results}
Although, the propagation of light through I-AFC and AFC is identical, in physical scenario the efficiencies can be very different for the two. This is due to the fact that in AFC a uniform frequency comb can be constructed on demand and as desired; however, in I-AFC the shape of the comb is determined by the atomic structure. In general, naturally available frequency combs are not uniform. Moreover, the fluctuations in the applied magnetic field and spatial distribution of the atoms in the ensemble may also result in non-uniform frequency comb. There are several other factors which can cause low efficiency in I-AFC. In this section, we address the effects of (i) random and non-uniform comb spacing, (ii) random and non-uniform optical depth on the efficiency of quantum memory.

The random effects are incorporated in the macroscopic polarization by averaging over the fluctuating polarization with a random parameter $\zeta$ which occurs with probability distribution $P(\zeta)$. The formal expression for the average polarization reads
\begin{align}
  \mathcal{P}( \omega)  = \tilde{\mathcal E}( \omega) \int  \mathcal{D}( \omega,\zeta)  ~P(\zeta) ~ d\zeta. \label{avgp}
\end{align}
For the purpose of calculations, we consider an I-AFC with a total of seven teeth at temperature $T = 10$ K and  light pulse with Gaussian spectrum of the following form:
\begin{align}
\mathcal{\tilde{E}}(0,\omega)= \exp[-\dfrac{\omega^2}{2(2\Delta)^2}],
\end{align}
where $\Delta$ is the average comb spacing. The total spectral width of the photon   is chosen in such a way that it covers all the seven peaks of the frequency comb. 

\subsection{Random and Non-uniform comb spacing} \label{srs}

\begin{figure*}[!tbh]
\centering
\subfigure[\label{rs}]{\includegraphics[scale=0.7]{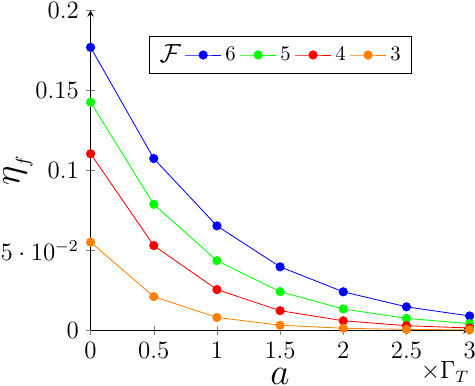}}\\
\subfigure[\label{rrs2}]{\includegraphics[scale=0.7]{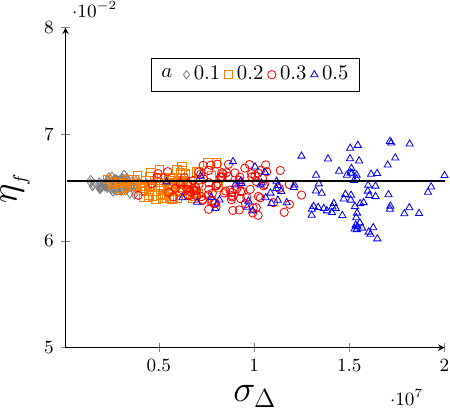}}
\subfigure[\label{rrs3}]{\includegraphics[scale=0.7]{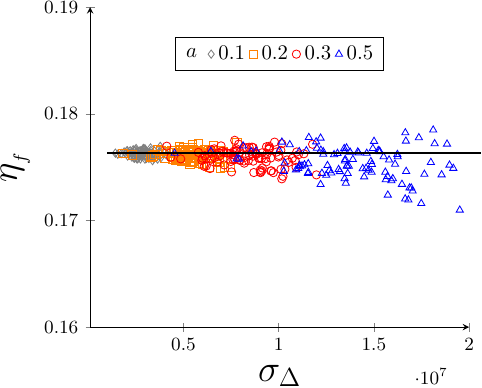}}
\subfigure[\label{cs1}]{\includegraphics[scale=0.7]{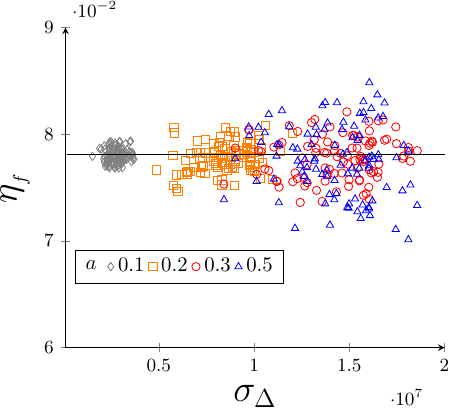}}
\subfigure[\label{rsf1}]{\includegraphics[scale=0.7]{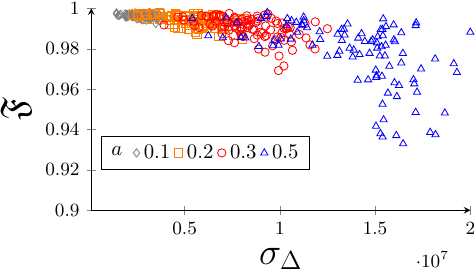}}
\subfigure[\label{rsf2}]{\includegraphics[scale=0.7]{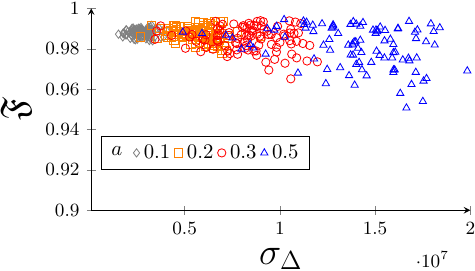}}
\subfigure[\label{cs2}]{\includegraphics[scale=0.7]{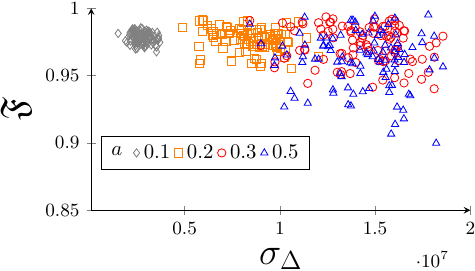}}
\caption{\subref{rs} Here we have plotted the efficiencies as a function of the fluctuation strength $a$ of the comb spacing and for various values of the finesse $\mathcal{F} = \Delta/(\gamma+2\Gamma_T)$. We have chosen the optical depth $\beta=\alpha L = 30$, tooth width $\gamma=5$MHz and the calculations are done at temperature $T = 10$K. From here we can see that although the efficiency is decreasing as we increase the randomness, the efficiencies are significant even at very large fluctuations. In \subref{rrs2}  and \subref{rrs3} we plot the efficiency for non-uniform comb with no fluctuations in the tooth-spacing. Here the non-uniformity is quantified by the standard deviation $\sigma_\Delta$ of mean frequency of each of the tooth in the frequency comb. We have set $\mathcal{F}=3$,~$\beta=18$ for Fig.~\subref{rrs2} and $\mathcal{F}=6$,~$\beta=30$ for Fig.~\subref{rrs3}, and the dashed line represents efficiency for the uniform comb. In Fig.~\subref{cs1} we plot efficiency of Cesium atom I-AFC at magnetic field strength $B=0.1T$.
In \subref{rsf1}, \subref{rsf2} and \subref{cd1} we plot the corresponding fidelities for the figures \subref{rrs2}, \subref{rrs3} and \subref{cs1}} 
\end{figure*}

\begin{figure*}[!htb]
\centering
\subfigure[\label{od2}]{\includegraphics[scale=0.7]{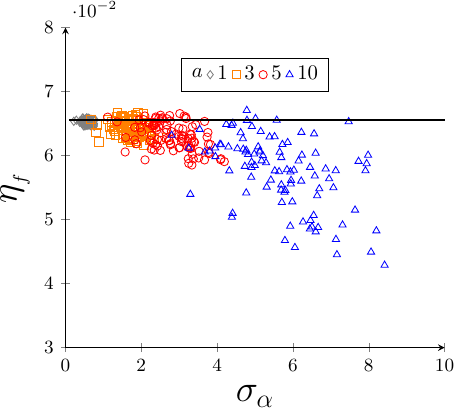}}
\subfigure[\label{od3}]{\includegraphics[scale=0.7]{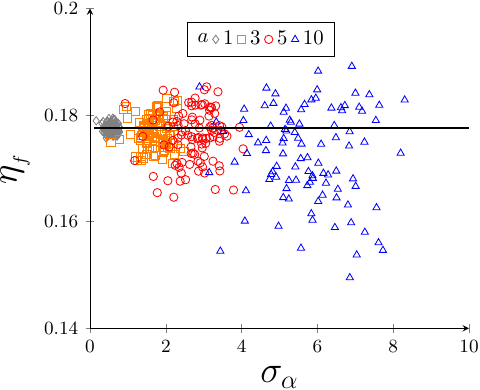}}
\subfigure[\label{cd1}]{\includegraphics[scale=0.7]{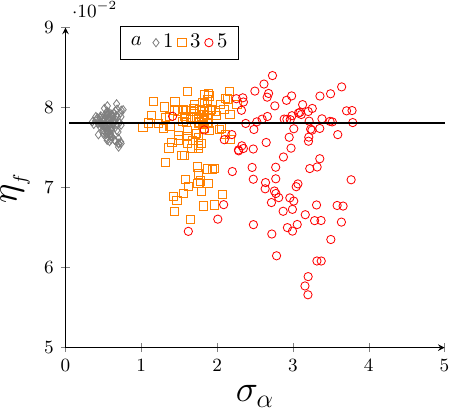}}
\subfigure[\label{fod1}]{\includegraphics[scale=0.7]{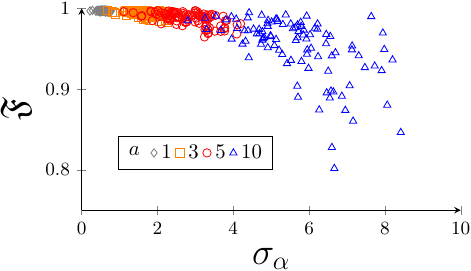}}
\subfigure[\label{fod2}]{\includegraphics[scale=0.7]{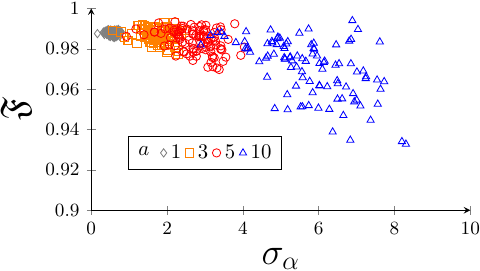}}
\subfigure[\label{cd2}]{\includegraphics[scale=0.7]{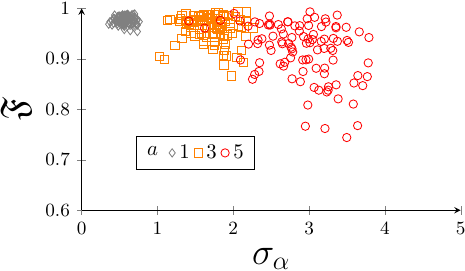}}
\caption{(Color online) In \subref{od2}  ($\mathcal{F}=3$,~$\beta=18$), \subref{od3} ($\mathcal{F}=6$,~$\beta=30$) and \subref{cd1} (Cesium at $0.1$T) we plot efficiency and standard deviation of optical depths over 100 trails where dashed line represents efficiency with zero randomness. In \subref{fod1}, \subref{fod2} and \subref{cd2} we polt the corresponding  fidelities.} \label{frod}
\end{figure*}

 Although, the spacing between different hyperfine energy levels is predominately determined by the applied external magnetic field, the spin-orbit coupling and interactions with the nuclear spins can result in non-uniform spacing between these energy levels. This non-uniformity can make a frequency comb unusable for purpose of quantum memory. Furthermore, when we dope these atoms in some dielectrics, or due to stray electric and magnetic fields, there can be fluctuations in the mean frequencies of different transitions. In this section, we address the effect of these adversaries on the efficiency of the I-AFC based quantum memory.

We start with fluctuations in the comb spacing, which is  introduced by shifting the comb spacing by a random number $\zeta$ with probability $P(\zeta) \propto \exp[-\zeta^2/(2\sigma_\zeta^2)]$. Here we have assumed the probability distribution to be Gaussian with width $\sigma_\zeta$. In this scenario, the propagator $\mathcal{D}$ reads
\begin{align}
\mathcal{D}( \omega,\zeta) = \sum_{n=-3}^{3} \dfrac{\gamma\alpha_n}{\gamma/2 +\Gamma_T+ \mi(n\Delta+ \zeta_n  +\omega)}+\dfrac{\mi \omega}{c}.\label{rse}
\end{align}
Following the approach given in Eq.~\eqref{ap} the averaged propagator is written as
\begin{align}
\mathcal{D}( \omega) = \sum_{n=-3}^{3} \dfrac{\gamma\alpha_n}{\gamma/2 +\Gamma_T+ a+\mi(n\Delta+ \omega)}+\dfrac{\mi \omega}{c}, \label{aps}
\end{align}
where $2a\propto \sigma_\zeta$ is the FWHM of the probability distribution. Eq.~\eqref{aps} shows that a general randomness in the comb spacing increases the broadening similar to the Doppler broadening. In Fig.~\ref{rs} we plot the forward  efficiencies as a function of width $a$ for different values of finesse $\mathcal{F}$. Here $\Gamma_T ~(\sim 5 \times 10^7) \gg\gamma ~(\sim 5 \times 10^6)$ thus making the  Doppler broadening dominant over natural broadening and we take the randomness $a$ as a multiple of the Doppler width $\Gamma_T$. We observe that the efficiency drops with increasing randomness. However, increasing the  finesse results in higher efficiency; hence, the effect of the fluctuating comb spacing can be mitigated by applying stronger magnetic field.

Next, we consider the case when the frequency comb is not uniform but there is no fluctuations in the mean frequencies. For such systems the propagator $\mathcal{D}$ can be written as 
\begin{align}
\mathcal{D} = \sum_{n=-3}^{3} \dfrac{\gamma\alpha_n}{\gamma/2 +\Gamma_T+\mi(\delta_n+ \omega)}+\dfrac{\mi \omega}{c}. 
\end{align}
For simplicity, we can assume the frequency $\delta_n$ of each of the tooth as a deviation from a average position, i.e., $\delta_n = n\Delta + \zeta_n$, where $\zeta_n$ is fixed and sampled randomly from the set $[-a\Gamma_T,a\Gamma_T]$. We use the standard deviation in the mean frequency of each of the tooth of the frequency comb as the measure for the non-uniformity in the frequency comb. The standard deviation can be calculated using the following expression
\begin{align}
\sigma^2_\Delta=\frac{1}{N-1} \sum_{n=-3}^3  {\zeta^2_n}-\expval{\zeta}^2,
\end{align}
where $N$ is the number of teeth.

In Figs.~\ref{rrs2} and~\ref{rrs3} we plot the numerically optimized efficiencies as a function of standard deviation in comb spacing for different finesse. Since multiple non-uniform combs can have the same value of the standard deviation $\sigma_\Delta$, we can have a band of efficiencies for the  same $\sigma_\Delta$.  We see that this band broadens with the standard deviation $\sigma_\Delta$ which itself is a function of $a$. An interesting finding of this plot is that occasionally large $\sigma_\Delta$ can result in higher efficiencies than the perfectly uniform combs. In other words, the uniform combs are not optimum for efficient quantum memory. For example in Cesium atom at temperature $10K$ and magnetic field $B=0.1T$ gives maximum efficency of $\eta_f=0.85$ and we can also notice that Cesium atom at $0.1T$ is similar to I-AFC with the parameters $\mathcal{F}=3$ and $\beta=18$. 

Although there is no obvious correlation between optimized efficiency and the comb spacing we observed that efficiencies are always higher than the uniform comb  when a random stretching with zero randomness in the cental peak is introduced i.e. $\zeta_n=[\zeta_{-3},\zeta_{-2},\zeta_{-1},0,\zeta_{1},\zeta_{2},\zeta_{3}]$ where $\zeta_{\mp n}$ are random negative and positive numbers
and similarly a random compression i.e. $\zeta_n=[\zeta_{3},\zeta_{2},\zeta_{1},0,\zeta_{-1},\zeta_{-2},\zeta_{-3}]$ resulted in lower efficiencies.

\subsection{Random and non-uniform optical depths}\label{Sec:OpticalDepth}

The optical depth of various teeth in I-AFC in natural atoms is non-uniform in general. The non-equal transition probability between  different atomic states is one of the biggest contributor to such non-uniformity. This can be further exaggerated by random environmental effects. In this section we study the effect of non-uniform optical depth and fluctuation therein of different teeth of the I-AFC on the efficiency of the quantum memory.  Similar to Sec.~\ref{rs}, we incorporate the fluctuating optical depth in I-AFC dynamics by adding randomness in the propagator $\mathcal{D}$ as follows
\begin{align}
\mathcal{D} = \sum_{n} \dfrac{(\gamma\alpha_n+\zeta_n)}{\gamma/2 +\Gamma_T+ \mi (n \Delta+\omega)}+\dfrac{\mi \omega}{c},\label{rop}
\end{align}
where $\zeta_n$ determine the fluctuations in the optical depth and it occurs with probability $P(\zeta) \propto \exp[-\zeta^2/2\sigma_\zeta^2]$. The average propagator can be calculated by taking average of Eq.~\eqref{rop} with the probability function $P(\zeta)$. Since the probability is an even function of $\zeta$, from Eq.~\eqref{avgp} we can see that the randomness has no effect on the optical depths. This shows that the fluctuating optical depths do not affect the efficiency of the quantum memory. 

However, the non-uniform optical depth without fluctuations can result in the loss of efficiency. Therefore, the propagator for such systems is defined by  Eq.~\eqref{rop} but there is no probability distribution over $\zeta$.
But  $\zeta_n$ is sampled from the set $[-a,a]$. The non-uniformity of the comb is quantified by the  standard deviation which is defined as $ \sigma^2_\alpha=1/(N-1) \sum_{n=-3}^3  {\zeta^2_n}-\expval{\zeta}^2 $ for $N$ number of teeth.

In Fig.~\ref{frod} we plot efficiency as a function of standard deviation in optical depth for different finesse. The calculations are done at temperature $T=10$K. The results for non-uniform optical depth are qualitatively similar to the one we got for non-uniform comb spacing. In this case also we see the non-uniform comb occasionally yield more efficient quantum memory as oppose to uniform comb. Although Cesium atom is similar to I-AFC $(\mathcal{F}=3,\beta=18)$ the echo is spoiled for high randomness $(a\geq 10)$. Unlike the stretching and compression of comb spacing, no correlation was observed between the efficiency and the randomness in optical depths.

\section{ Conclusion}\label{Sec:Conclusion}
In conclusion, we have shown that the I-AFC based quantum memory is robust against non-uniformity in the comb spacing, and non-uniform and fluctuating optical depths. The fluctuations in the comb spacing can affect the quantum memory efficiency in a strong way. However, this effect can be easily mitigated by increasing the finesse of the I-AFC, which can be done by applying stronger magnetic field.
Our study shows that even the imperfect atomic systems in the extreme environmental conditions can be used for efficient I-AFC based quantum memories highlighting the robust nature of I-AFC based quantum memory.

\section{Acknowledgements}
We acknowledge the financial support from Inter-disciplinary Cyber Physical Systems(ICPS) programme of the Department of Science and Technology, India, (Grant No.:DST/ICPS/QuST/Theme-1/2019/12)


\end{document}